\documentclass[eqsecnum,twocolumn,amsmath,showpacs,amsfonts,aps,prc,floatfix]{revtex4}
\usepackage{graphicx}
\usepackage{bm}

\begin{document}

\title{Causal dissipative hydrodynamics for QGP fluid in 2+1 dimensions}

\author{A. K. Chaudhuri}
\email[E-mail:]{akc@veccal.ernet.in}
\affiliation{Variable Energy Cyclotron Centre, 1/AF, Bidhan Nagar, 
Kolkata 700~064, India}

\begin{abstract}
In 2nd order causal dissipative theory, space-time evolution of QGP fluid is 
studied in 2+1 dimensions.  Relaxation equations for shear stress tensors are
solved simultaneously with the energy-momentum conservation equations. Comparison 
of evolution of ideal and viscous QGP fluid, initialized under the same conditions, e.g. same equilibration time, energy density and velocity profile, indicate that 
in a viscous dynamics, energy density or temperature of the fluid evolve slowly, than in an ideal fluid. Cooling gets slower as viscosity increases. Transverse expansion  also increases in a viscous dynamics. For the first time we have also studied elliptic flow of 'quarks' in causal viscous dynamics. It is shown that elliptic flow of quarks saturates due to non-equilibrium correction to equilibrium distribution function,
and can not be mimicked by an ideal hydrodynamics. 
\end{abstract}


\date{\today}  

\maketitle

\section{Introduction}
\label{sec1}
%
One of the most important discoveries in Relativistic Heavy ion collider (RHIC) at Brokhaven National Laboratory is the large elliptic flow in non-central Au+Au collisions \cite{BRAHMSwhitepaper,PHOBOSwhitepaper,PHENIXwhitepaper,STARwhitepaper} . Elliptic flow measures the momentum anisotropy of produced
particles and is quantified by the 2nd harmonic of the azimuthal distribution,

\begin{equation}
v_2(p_T)=<cos(2\phi)>=
\frac
{\int_0^{2\pi} \frac{dN}{dyd^2p_T} \cos(2\phi) d\phi}
{\int_0^{2\pi} \frac{dN}{dyd^2p_T}  d\phi}
\end{equation}

Elliptic flow is naturally explained in hydrodynamics.
Hydrodynamic pressure is built up from rescattering of 
secondaries, and pressure gradients drive the subsequent collective motion. 
In non-central Au+Au collisions, initially, the reaction zone is 
asymmetric (almond shaped). The pressure gradient is large in one direction and small in the other. The asymmetric pressure gradients generates the elliptic flow. Naturally, in a central collision, reaction zone is symmetric and elliptic flow vanishes. Observed elliptic flow then give the strongest indication that in non-central  Au+Au collisions, a collective QCD matter is produced. 
Whether the formed matter can be identified as   the much sought after Quark-Gluon Plasma (QGP) as predicted in Lattice QCD simulations \cite{lattice} is presently debatable. 

Ideal hydrodynamics has been partly successful in explaining the
observed elliptic flow, quantitatively \cite{QGP3}. 
Elliptic flow of identified particles, up to $p_T\sim$1.5 GeV are
well reproduced in ideal hydrodynamics.
Ideal hydrodynamics also explains the transverse momentum spectra of identified particles (up to $p_T\sim$ 1.5 GeV).
Success of {\em ideal} hydrodynamics 
in explaining bulk of the data \cite{QGP3}, together with the string theory motivated lower limit of shear viscosity $\eta/s \geq 1/4\pi$ \cite{Policastro:2001yc,Policastro:2002se} has led to a paradigm that in Au+Au collisions, a nearly perfect fluid is created.  

However, the paradigm of "perfect fluid" produced in Au+Au collisions at RHIC need to be clarified. As indicated above, the ideal hydrodynamics is only partially successful and in a limited $p_T$ range ($p_T \leq$1.5 GeV) \cite{Heinz:2004ar}. The transverse momentum spectra 
of identified particles also starts to deviate form ideal fluid dynamics prediction beyond $p_T\approx$ 1.5 GeV. Experimentally determined  HBT radii are not reproduced in the ideal fluid dynamic models, the famous "HBT puzzle" \cite{Heinz:2002un}.  
It also do not reproduce the experimental trend that elliptic flow saturates at large transverse momentum.   
These shortcomings of ideal fluid dynamics indicate
greater importance of dissipative effects  in the $p_T$ ranges greater than 1.5 GeV or in more peripheral collisions. Indeed, 
ideal fluid is a concept, never realized in nature. 
As suggested
in string theory motivated models \cite{Policastro:2001yc,Policastro:2002se},  QGP viscosity could be small,   $\eta/s \geq 1/4\pi$, nevertheless it is non-zero.   
It is important
to study the effect of viscosity, even if small, on space-time evolution of QGP fluid and quantify its effect. This requires a numerical implementation of relativistic dissipative fluid dynamics. Furthermore, 
if QGP fluid is formed in heavy ion collisions, it has to be characterized by measuring its transport coefficients, e.g. heat conductivity, bulk and shear viscosity. 
Theoretically, it is possible to obtain those transport coefficients in a kinetic theory model. However,
in the present status of theory,  the goal can not be achieved immediately, even more so for a strongly interacting QGP (sQGP).
Alternatively, one can use the
experimental data to obtain a "phenomenological" limit of transport coefficients of sQGP. 
It will also require a numerical implementation of relativistic dissipative fluid dynamics. There is another incentive to study 
dissipative   hydrodynamics. Ideal hydrodynamics depends on the
assumption of local equilibrium. Before local equilibrium is attained, the system has to pass through a non-equilibrium stage,
where (if non-equilibrium effects are small) dissipative hydrodynamics may be applicable.  Indeed, we can explore early times of fluid evolution better in a dissipative hydrodynamics. 

Theory of dissipative relativistic 
fluid has been formulated quite early.  The original dissipative relativistic fluid equations were given by Eckart \cite{Eckart}
and Landau and Lifshitz \cite{LL63}. They are called 1st order theories. Formally, relativistic dissipative hydrodynamics are obtained from an expansion of entropy 4-current, in terms of dissipative fluxes. In 1st order theories, entropy 4-current contains terms linear in dissipative quantities.   
1st order theory of dissipative hydrodynamics 
suffer from the problem of causality violation. Signal can travel faster than light.  Causality violation is unwarranted in any theory,
even more in a relativistic theory. 
 The problem of causality violation is removed in the Israel-Stewart's 2nd order theory of dissipative fluid \cite{IS79}. In 2nd order theory, 
expansion of entropy 4-current contains terms 2nd order in dissipative fluxes. However, these leads to complications that
dissipative fluxes are no longer function of the state variables only.
They become dynamic.
The space of  thermodynamic variables has to be extended to include the dissipative fluxes (e.g. heat conductivity, bulk and shear viscosity).

Even though 2nd order theory was formulated some 30 years back, significant progress towards its numerical implementation 
has only been made very recently \cite{Muronga:2001zk,Teaney:2004qa,MR04,Heinz:2005bw,Chaudhuri:2005ea,Chaudhuri:2006jd,Chaudhuri:2007yn,Chaudhuri:2007yk,Koide:2007kw,Baier:2006gy}. At the Cyclotron Centre, Kolkata, we have developed a code  "AZHYDRO-KOLKATA"  to simulate the  
hydrodynamic evolution of dissipative QGP fluid. Presently only dissipative effect included is the shear viscosity.   
Some results of AZHYDRO-KOLKATA,  for first order dissipative hydrodynamics have been published earlier \cite{Chaudhuri:2006jd,Chaudhuri:2007yn,Chaudhuri:2007yk}.
In the present paper, for the first time, we will present some results    for 2nd order dissipative hydrodynamics
in 2+1 dimensions. 
In the present paper, we will consider effect of dissipation in the QGP phase only.
Effect of phase transition will be studied in a later publication. 

The paper is organized as follows: In section~\ref{sec2} we briefly
review relativistic dissipative fluid dynamics.
In section~\ref{sec3} we derive the relevant equations in 2+1 dimension (assuming boost-invariance). Required inputs e.g. the equation of state, viscosity coefficient and initial conditions are discussed in section ~\ref{sec4}. Simulation results
from AZHYDRO-KOLKATA are shown in section ~\ref{sec5}.  
In section \ref{sec6} we compare the transverse momentum spectra and elliptic flow  of quarks in ideal and viscous dynamics. The concluding section~\ref{sec7} summarizes our results.

\section{dissipative fluid dynamics}
\label{sec2}
%

In this section, I briefly discuss the phenomenological theory of
dissipative hydrodynamics. More detailed exposition can be found in \cite{IS79}.  

A simple fluid, in an arbitrary state, is fully specified by primary variables: particle current ($N^\mu$), energy-momentum tensor ($T^{\mu\nu}$) and
entropy current ($S^\mu$) and a number of additional (unknown) variables. Primary variables satisfies the conservation
laws;

\begin{eqnarray} 
\partial_\mu N^\mu =&&0,\label{eq1}\\
\partial_\mu T^{\mu\nu}=&&0, \label{eq2}
\end{eqnarray} 

\noindent and the 2nd law of thermodynamics,

\begin{equation}
\partial_\mu S^\mu  \geq 0. \label{eq3}
\end{equation}

In relativistic fluid dynamics, one defines a time-like hydrodynamic 4-velocity,
$u^\mu$ (normalized as $u^2=1$). One also define 
a projector, 
$\Delta^{\mu\nu}=g^{\mu\nu}-u^\mu u^\nu$,
orthogonal to the 4-velocity ($\Delta^{\mu\nu}u_\nu=0$).
In equilibrium, an unique
4-velocity ($u^\mu$) exists such that the particle density ($n$), energy density ($\varepsilon$) and the entropy density ($s$) can be obtained from,

\begin{eqnarray} \label{eq4}
 N^\mu_{eq}=&& n u_\mu  \\
\label{eq5}
T^{\mu\nu}_{eq}=&&\varepsilon u^\mu u^\nu -p \Delta^{\mu\nu}\\
\label{eq6}
S^\mu_{eq}=&&s u_\mu 
\end{eqnarray}

An equilibrium state is assumed to be fully specified by 5-parameters,
$(n,\varepsilon,u^\mu)$ or equivalently by the thermal potential,
$\alpha=\mu/T$ ($\mu$ being the chemical potential) and inverse 4-temperature, $\beta^\mu=u^\mu/T$. Given a equation of state, $s=s(\varepsilon,n)$, pressure $p$ can be obtained from the generalized thermodynamic relation,

\begin{equation} \label{eq7}
S^\mu_{eq}=p\beta^\mu-\alpha N^\mu_{eq} +\beta_\lambda T^{\lambda\mu}_{eq}
\end{equation} 

Using the Gibbs-Duhem relation, 
$d(p\beta^\mu)=N^\mu_{eq} d\alpha -T^{\lambda\mu}_{eq}d\beta_\lambda$, following relations can be established on the equilibrium hyper-surface $\Sigma_{eq}(\alpha,\beta^\mu)$,


\begin{equation} \label{eq8}
dS^\mu_{eq}=-\alpha dN^\mu_{eq}+\beta_\lambda dT^{\lambda\mu}_{eq}
\end{equation}

In a non-equilibrium system, no 4-velocity can be found such that Eqs.\ref{eq4},\ref{eq5},\ref{eq6} remain valid. Tensor decomposition leads to additional terms,  

\begin{eqnarray}\label{eq9}
N^\mu=&&N^\mu_{eq}+\delta N^\mu=nu^\mu + V^\mu\\
\label{eq10}
T^{\mu\nu} =&&T^{\mu\nu}_{eq}+\delta T^{\mu\nu} \nonumber \\
=&&
[\varepsilon u^\mu u^\nu-p\Delta^{\mu\nu}]+\Pi\Delta^{\mu\nu} + \pi^{\mu\nu} \nonumber\\
&&+(W^\mu u^\nu + W^\nu u^\mu)\\
\label{eq11}
S^\mu=&&S^\mu_{eq}+\delta S^\mu=su^\mu + \Phi^\mu
\end{eqnarray} 

The new terms describe a net flow of charge $V^\mu=\Delta^{\mu\nu} N_\nu$, heat flow, $W^\mu=(\varepsilon+p)/n V^\mu +q^\mu$ (where $q^\mu$ is the heat flow vector), and entropy flow $\Phi^\mu$. 
$\Pi=-\frac{1}{3}\Delta_{\mu\nu}T^{\mu\nu}-p$ is the bulk viscous pressure
and $\pi^{\mu\nu}= [\frac{1}{2}(\Delta^{\mu\sigma}\Delta^{\nu\tau}+ 
\Delta^{\nu\sigma}\Delta^{\mu\tau}-\frac{1}{3}
\Delta^{\mu\nu}\Delta^{\sigma\tau}]T_{\sigma\tau}$ is the shear stress tensor.
Hydrodynamic 4-velocity can be chosen
to eliminate either $V^\mu$ (the Eckart frame, $u^\mu$ is parallel 
to particle flow) or the heat flow $q^\mu$ (the Landau frame, $u^\mu$ is
parallel to energy flow). In relativistic heavy ion collisions,
central rapidity region is nearly baryon free and Landau's frame is more appropriate than the Eckart's frame. Dissipative flows are transverse to $u^\mu$ and additionally, shear stress tensor is traceless. Thus a non-equilibrium state  require 1+3+5=9 additional quantities, the dissipative 
flows $\Pi$, $q^\mu$ (or $V^\mu$) and $\pi^{\mu\nu}$.  
In kinetic theory, $N^\mu$ and $T^{\mu\nu}$ are the 1st and 2nd moment of the distribution function. Unless the function is known a-priori, two moments do not furnish enough information to enumerate the microscopic states required to determine $S^\mu$, and
in an arbitrary non-equilibrium state, no relation exists between,
$N^\nu$, $T^{\mu\nu}$ and $S^\mu$. 
{\em Only in a state, close to 
a equilibrium one, such a relation can be established}. 
Assuming that the equilibrium relation Eq.\ref{eq8} remains valid
in a "near equilibrium state" also, the entropy current can be generalized as,

\begin{equation} \label{eq12}
S^\mu=S^\mu_{eq}+dS^\mu
=p\beta^\mu-\alpha N^\mu +\beta_\lambda T^{\lambda\mu} + Q^\mu
\end{equation}

\noindent where $Q^\mu$ is an undetermined quantity in 2nd order in deviations, $\delta N^\mu=N^\mu-N^\mu_{eq}$ and $\delta T^{\mu\nu}=T^{\mu\nu}-T^{\mu\nu}_{eq}$.  Detail form of $Q^\mu$ is constrained by the 2nd law $\partial_\mu S^\mu \geq 0$.
With the help of conservation laws and Gibbs-Duhem relation,
entropy production rate can be written as,

\begin{eqnarray} \label{eq13}
\partial_\mu S^\mu=-\delta N^\mu \partial_\mu \alpha
+\delta T^{\mu\nu} \partial_\mu \beta_\nu + \partial_\mu Q^\mu
\end{eqnarray}

Choice of $Q^\mu$ leads to 1st order or 2nd order theories of dissipative hydrodynamics.
In 1st order theories the simplest choice is made, $Q^\mu=0$, entropy current contains terms up to 1st order in deviations,
$\delta N^\mu$ and $\delta T^{\mu\nu}$. Entropy production rate can be written as,

\begin{equation}\label{eq14}
T\partial_\mu S^\mu
=\Pi X -q^\mu X_\mu + \pi^{\mu\nu} X_{\mu\nu}  
\end{equation}

\noindent where, $X=-\nabla.u$; $X^\mu=\frac{\nabla^\mu}{T}-u^\nu \partial_\nu u^\mu$ 
and 
$X^{\mu\nu}=\nabla^{<\mu} u^{\nu>}$.
   
The 2nd law, $\partial_\mu S^\mu \geq 0$ can be satisfied by postulating a linear relation between the dissipative flows and thermodynamic forces,
 
\begin{eqnarray}
\label{eq15}
\Pi=&&-\zeta \theta,\\
\label{eq16}
q^\mu=&&-\lambda \frac{nT^2}{\varepsilon+p}\nabla^\mu(\mu/T),\\ 
\label{17}
\pi^{\mu\nu}=&&2\eta \nabla^{<\mu}u^{\nu>}
\end{eqnarray}

\noindent where $\zeta$, $\lambda$ and $\eta$ are the positive transport coefficients, bulk viscosity, heat conductivity and shear viscosity respectively. 

In 1st order theories, causality is violated. If, in a given fluid cell, at a certain time, thermodynamic forces vanish, corresponding dissipative fluxes also vanish instantly. Violation of causality is unwanted in any theory, even more so in relativistic theory. Causality violation of dissipative hydrodynamics is corrected in   2nd order theories \cite{IS79}. In 2nd order theories, entropy current contain terms up to 2nd order in the deviations, $Q^\mu \neq 0$. 
The most general $Q^\mu$ containing terms up to 2nd order in deviations can be written as,

\begin{equation} \label{eq18}
Q^\mu=-(\beta_0\Pi^2-\beta_1 q^\nu q_\nu + \beta_2\pi_{\nu\lambda}\pi^{\nu\lambda})
\frac{u^\mu}{2T} -\frac{\alpha_0\Pi q^\mu}{T} +\frac{\alpha_1 \pi^{\mu\nu}q_\nu}{T}
\end{equation}

As before, one can cast the entropy production rate ($T\partial_\mu S^\mu$) in the form of Eq.\ref{eq14}. 
Neglecting the terms involving dissipative flows with gradients of equilibrium thermodynamic quantities (both are assumed to be small) and demanding that a linear
relation exists between the dissipative flows and thermodynamic
forces,  following {\em relaxation} equations for the dissipative flows can be obtained,  

\begin{eqnarray} \label{eq19}
\Pi=&&-\zeta (\theta +\beta_0 D\Pi)\\
\label{eq20}
q^\mu=&&-\lambda \left[ \frac{nT^2}{\varepsilon+p}\nabla^\mu(\frac{\mu}{T}) 
-\beta_1 Dq^\mu \right]\\
\label{eq21}
\pi^{\mu\nu}=&&2\eta \left[\nabla^{<\mu}u^{\nu>} -\beta_2 D\pi_{\mu\nu} \right],
\end{eqnarray}

\noindent where  $D=u^\mu \partial_\mu$is the convective time derivative. Unlike in the 1st order theories, in 2nd order theories,
dynamical equations control the dissipative flows.  
Even if thermodynamic forces vanish, dissipative flows do not vanish instantly.  

Before we proceed further, it may be mentioned that the parameters,
$\alpha$ and $\beta_\lambda$ are not connected to the actual state
($N^\mu,T^{\mu\nu}$). The pressure $p$ in Eq.\ref{eq12} is also not
the "actual" thermodynamics pressure, i.e. not the work done in
an isentropic expansion. Chemical potential $\alpha$ and 4-inverse temperature $\beta_\lambda$ has meaning only for the equilibrium state. Their meaning need not be extended to non-equilibrium states also. However, it is possible to fit a fictitious "local equilibrium" state, point by point, such that pressure $p$ in
Eq.\ref{eq12} can be identified with the thermodynamic pressure,  
at least up to  1st order. The conditions of fit fixes the  
underlying non-equilibrium phase-space distribution.

\section{(2+1)-dimensional viscous hydrodynamics with
longitudinal boost invariance}
\label{sec3}
Complete dissipative hydrodynamics is a numerically challenging problem. It requires simultaneous solution of 14 partial differential equations (5 conservation equations and 9 relaxation equations for dissipative flows). We reduce the problem to solution of 6 partial differential equations 
(3 conservation equations and 3 relaxation equations).
In the following, we will study boost-invariant  evolution of baryon free QGP fluid, including the dissipative effect due to shear viscosity only. Shear viscosity is the most important dissipative effect.  
For example, in  a baryon free QGP, heat conduction is zero and we can disregard Eq.\ref{eq20}. Bulk viscosity is also zero for the QGP fluid (point particles) and Eq.\ref{eq19}
can also be neglected. Shear pressure tensor has 5 independent 
components but the assumption of boost invariance reduces the
number of independent components to three. 
For a baryon free fluid, we can also disregard the conservation equation Eq.\ref{eq1}.  With the assumption of boost-invariance,
 energy-momentum conservation equation  $\partial_\mu T^{\mu \eta}=0$  become redundant and  
only three energy-momentum conservation equations are required to be solved.  

\begin{figure}[ht]
\includegraphics[bb=14 13 581 829,width=0.9\linewidth,clip]{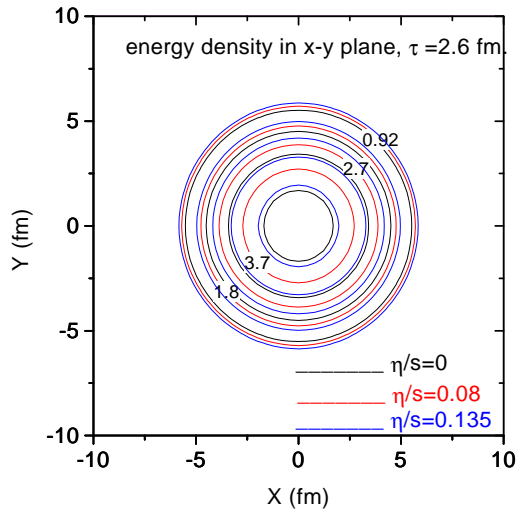}
\vspace{-4cm}
\caption{(color online). Constant energy density contours in x-y plane at 
$\tau$=2.6 fm. The black  lines are for ideal fluid ($\eta/s$=0). The 
red and blue lines are for viscous fluid with ADS/CFT and perturbative estimate of viscosity, $\eta/s$=0.08 and 0.135.}
\label{F1}
\end{figure}

Heavy ion collisions
are best described in ($\tau,x,y,\eta$) coordinates, where $\tau=\sqrt{t^2-z^2}$ is the longitudinal proper time and  $\eta=\frac{1}{2} \ln \frac{t+z}{t-z}$ is the space-time rapidity.
$r_\perp=(x,y)$ are the usual cartisan coordinate in the plane, 
transverse to the beam direction. Relevant equations concerning this coordinate transformations are given in the appendix \ref{app1}.

Explicit equations for energy-momentum conservation
in ($\tau$,x,y,$\eta$) coordinates are given in the appendix \ref{app2}.
 We note that unlike in ideal fluid, in viscous fluid dynamics, conservation equations (see Eqs.\ref{eqb1}-\ref{eqb3}) contain additional pressure gradients due to shear viscosity.  Both $T^{\tau x}$ and $T^{\tau y}$ components of energy-momentum tensor now evolve under additional pressure gradients. The rightmost term of Eq.\ref{eqb3} also indicate that in viscous dynamics, longitudinal pressure  is 
effectively reduced (note that the $\pi^{\eta\eta}$ component is negative). Since pressure can not be negative, shear viscosity is limited by the condition, $p + \tau^2\pi^{\eta\eta} \geq 0$.

As evident from the Eqs.\ref{eqb1}-\ref{eqb3}, in boost-invariant dissipative hydrodynamics, with shear viscosity taken into account, fluid   evolution depends only on seven components of the shear stress tensors. They are $\pi^{\tau\tau}$, $\pi^{\tau x}$, $\pi^{\tau y}$, $\pi^{xx}$, $\pi^{yy}$, $\pi^{xy}$ and $\pi^{\eta\eta}$. However, all the seven components are not independent. Tracelessness, transversality to $u^\mu$ and the
assumption of boost-invariance reduces the independent components to  three. Presently, we choose $\pi^{xx}$ $\pi^{yy}$ and $\pi^{xy}$ as the independent components.  Relaxation equations for the independent components are given in the appendix \ref{app3} (see Eqs.\ref{eqc7}-\ref{eqc9}).
They are solved simultaneously with the three energy momentum conservation equations Eqs.\ref{eqb1}-\ref{eqb3}, with inputs as
discussed below.

\begin{figure}[ht]
\includegraphics[bb=14 13 581 829,width=0.9\linewidth,clip]{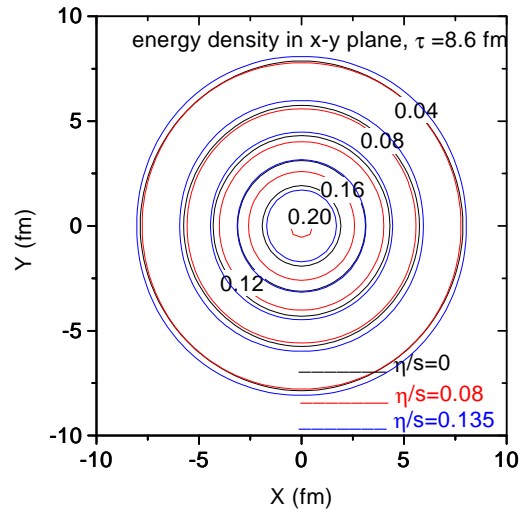}
\vspace{-4cm}
\caption{(color online). same as Fig.\ref{F1} but at time
$\tau$=8.6 fm.}
\label{F2}
\end{figure}

\section{Equation of state, viscosity coefficient and initial conditions}
\label{sec4}

\subsection{Equation of state}

One of the most important inputs of a hydrodynamic model is  
the equation of state. Through this input, the macroscopic hydrodynamic models make contact 
with the microscopic world. 
In the present demonstrative calculation we will show results for the QGP phase only. In the QGP phase, we use the simple equation of state,
$p=\frac{1}{3} \varepsilon$, with energy density given as,

\begin{equation}
\varepsilon=\frac{\pi^2}{30} g_{qgp} T^4
\end{equation}

\noindent where $g_{qgp}=g_{gluon}+\frac{7}{8}g_{quark}$ is the degeneracy factor for QGP. $g_{gluon}=2(helicity)\times 8(color)$ is the degeneracy factor for gluons and
$g_{quark}=2(spin)\times 3(color) \times 2(q+\bar{q})\times N_f$ is the degeneracy factor for $N_f$ flavored quarks. For $N_f\approx 2.5$, the degeneracy factor is $g_{qgp}=42.25$

\subsection{Shear viscosity coefficient}

Shear viscosity coefficient ($\eta$) of  QGP or sQGP
is quite uncertain. In a strongly coupled QGP, shear viscosity can not be computed.  Recently, using the ADS/CFT correspondence
 \cite{Policastro:2001yc,Policastro:2002se}
shear viscosity of a strongly coupled gauze theory, N=4 SUSY YM,
has been evaluated, $\eta=\frac{\pi}{8}N^2_cT^3$ and the    entropy
is given by $s=\frac{\pi^2}{2}N^2_cT^3$. Thus in the strongly coupled field theory,

\begin{equation} \label{eq26}
\left ( \frac{\eta}{s} \right )_{ADS/CFT} = \frac{1}{4\pi}\approx0.08,
\end{equation}

Shear viscosity is quite uncertain in perturbative QCD also.
At high temperature, shear viscosity, in leading log, can be written as \cite{Arnold:2000dr,Baym:1990uj},

\begin{equation} \label{eq24}
\eta=\kappa  \frac{T^3}{g^4 \ln g^{-1}},
\end{equation}

\noindent where $g$ is the strong coupling constant. The
leading log shear viscosity coefficient $\kappa$ depend on the number of fermion flavors ($N_f$). For example, for two flavored QGP, $\kappa=86.47$ and $\kappa=106.7$ for a three flavored QGP.
With entropy density of QGP, $s=\frac{\pi^2}{15} g_{qgp} T^3$. For two flavored QGP and 
$\alpha_s \approx$0.5, the ratio of viscosity over the entropy, in the perturbative regime is estimated as,

\begin{equation} \label{eq25}
\left (\frac{\eta}{s} \right )_{pert} \approx 0.135,
\end{equation}

For lower $\alpha_s$ , perturbative estimation of $\eta/s$ could be even higher. 

\begin{figure}[ht]
\includegraphics[bb=14 13 581 829,width=0.9\linewidth,clip]{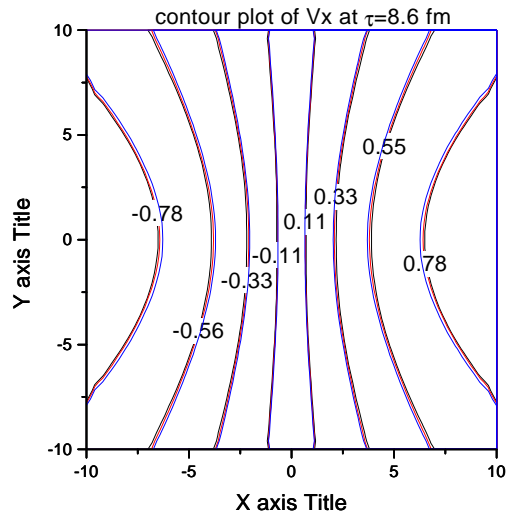}
\vspace{-4cm}
\caption{(color online). contours of constant $v_x$ in x-y plane at 
$\tau$=8.6 fm. The black  lines are for ideal fluid ($\eta/s$=0). The 
red and blue lines are for viscous fluid with ADS/CFT and perturbative estimate of viscosity, $\eta/s$=0.08 and 0.135.}
\label{F3}
\end{figure}

Shear viscosity can also be expressed in terms of sound attenuation
length, $\Gamma_s$, defined as,

\begin{equation} \label{eq27}
\Gamma_s=\frac{\frac{4\eta}{3}}{sT}
\end{equation}

$\Gamma_s$ is equivalent to mean free path and for a 
valid hydrodynamic description $\Gamma_s/\tau << 1$, i.e. mean
free path is much less than the system size.  
Initial conditions of the fluid must be chosen carefully such that
the validity condition $\Gamma_s/\tau << 1$ remains valid initially
as well as at later time also. In the present work, we have treated
viscosity as a parameter. To explore the effect of viscosity, we have used both the ADS/CFT estimate $\eta/s$=0.08 and
perturbative estimate $\eta/s$=0.135. We have also run the code 
with a higher value of viscosity $\eta/s$=0.2.  

\begin{figure}[ht]
\includegraphics[bb=14 13 581 829,width=0.9\linewidth,clip]{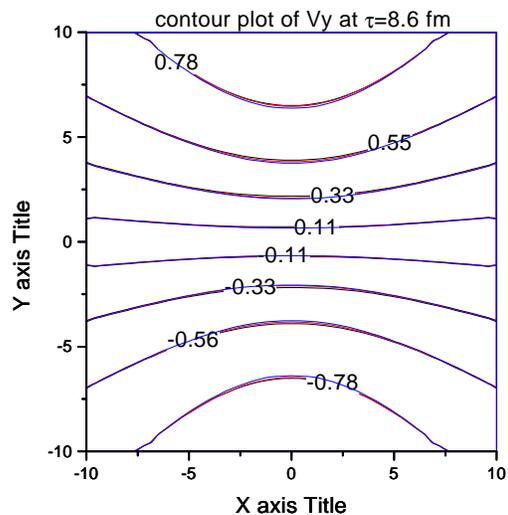}
\vspace{-4cm}
\caption{(color online). contours of constant $v_y$ in x-y plane at 
$\tau$=8.6 fm. The black  lines are for ideal fluid ($\eta/s$=0). The 
red and blue lines are for viscous fluid with ADS/CFT and perturbative estimate of viscosity, $\eta/s$=0.08 and 0.135.}
\label{F4}
\end{figure}

\subsection{Initial conditions}

Solution of Eqs.\ref{eqb1}-\ref{eqb3} 
require initial conditions, the initial time $\tau_i$, the transverse distribution of energy density $\varepsilon(x,y)$ and the velocities
$v_x(x,y)$ and $v_y(x,y)$.
Following \cite{QGP3}, initial
transverse energy density is parameterized geometrically.
At an impact parameter $\vec{b}$, transverse distribution of
wounded nucleons $N_{WN}(x,y,\vec{b})$    and of binary NN collisions 
$N_{BC}(x,y,\vec{b})$ to are calculated in a Glauber model. 
A collision at impact parameter $\vec{b}$ is assumed to contain 25\% hard scattering
(proportional to number of binary collisions) and 75\% soft scattering
(proportional to number of wounded nucleons).  
Transverse energy density 
profile at impact parameter $\vec{b}$ is then obtained as,

\begin{equation}
\varepsilon(x,y,\vec{b})=\varepsilon_0(0.75\times N_{WN}(x,y,\vec{b})+0.25 \times N_{BC}(x,y,\vec{b}))
\end{equation}

\noindent  with central energy density $\varepsilon_0$=30 $GeV/fm^{-3}$. The equilibration time is chosen as $\tau_i$=0.6 fm \cite{QGP3}. The initial velocities $v_x$ and $v_y$ are assumed to be zero initially.
\begin{figure}[ht]
\includegraphics[bb=14 13 581 829,width=0.9\linewidth,clip]{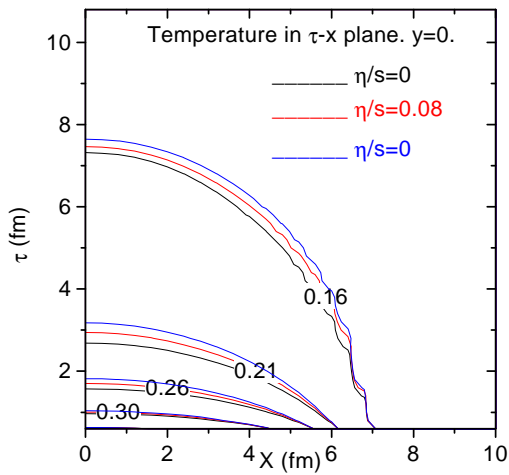}
\vspace{-4cm}
\caption{(color online). Constant temperature contour in $x-\tau$ plane, for fixed y=0. The black, red and blue lines are for ideal, viscous fluid with $\eta/s$=0.08 and viscous fluid with $\eta/s$=0.135.}
\label{F5}
\end{figure}

In dissipative hydrodynamics, one requires initial conditions for the viscous pressures also. Due to longitudinal boost invariance of the problem, we assume that  viscous pressures have attained their boost-invariant values at the time of equilibration. Boost invariant values of the three independent shear stress-tensors can be easily obtained from Eqs.\ref{eqc7}-\ref{eqc9}, $\sigma^{xx}=\sigma^{yy}=\theta=\frac{1}{\tau}$ and $\sigma^{xy}=0$ (  
at the initial time $\tau_i$, $u^\mu=(1,0,0,0)$, $Du^\mu=0$. The initial distribution of shear pressure tensors are then obtained
as,
 
\begin{eqnarray} 
\pi^{xx}(x,y,\vec{b})=&&2\eta \sigma^{xx}=2\eta/\tau_i\\ 
\pi^{yy}(x,y,\vec{b})=&&2\eta \sigma^{yy}=2\eta/\tau_i \\
\pi^{xy}(x,y,\vec{b})=&&2\eta\sigma^{xy}=0 
\end{eqnarray}

As stated earlier, 
the viscous coefficient $\eta$ is obtained using the relation,
$\eta /s =const$, $const$=0.08, 0.135 or 0.2.  
For these values of shear viscosity, the
 validity condition $\Gamma_s/\tau << 1$ is satisfied initially. The validity condition is better satisfied at later time.

\section{Results} \label{sec5}

\section{Stability of numerical solutions}

\subsection{Evolution of the viscous QGP fluid}

The energy-momentum conservation equations \ref{eqb1}-\ref{eqb3}, and the relaxation equations
\ref{eqc5}-\ref{eqc7}
 are solved simultaneously using the code, AZHYDRO-KOLKATA,  developed at the Cyclotron Centre, Kolkata. As mentioned earlier, we have solved the equations in the QGP phase only and did not consider any phase transition.
In the following we will show the results for central Au+Au  collisions (impact parameter $b=$ 0 fm). To understand the effect of shear viscosity, with the same initial conditions, we have solved the energy-momentum conservation equations for ideal fluid and viscous fluid.  As mentioned earlier, we have considered two values of viscosity, the ADS/CFT motivated value $\eta/s$=0.08 and the perturbative estimate, $\eta/s$=0.135. 

\begin{figure}[h]
\includegraphics[bb=14 13 581 829,width=0.9\linewidth,clip]{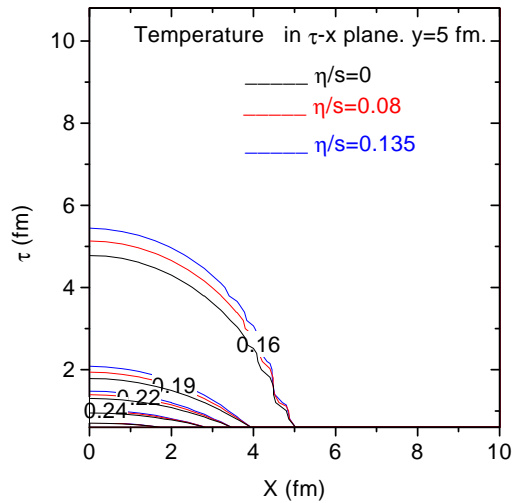}
\vspace{-4cm}
\caption{(color online). same as fig.\ref{F5} but at y=5.0}
\label{F6}
\end{figure}

In Fig.\ref{F1}, we have shown the contours of constant energy density in x-y plane, after an evolution of 2.6 fm.
The black lines are for ideal fluid evolution. The  red and blue lines are for 
viscous fluid with ADS/CFT ($\eta/s$=0.08) and perturbative ($\eta/s$=0.135) estimate of viscosity. Constant energy density
contours,
as depicted in Fig.\ref{F1}, indicate that with viscosity fluid cools slowly. Cooling
gets slower as viscosity increases.
Thus at any point in the x-y plane, energy density of viscous fluid is higher
than that of an ideal fluid.  
At later time also, compared to an ideal fluid, viscous fluid evolve slowly. In Fig.\ref{F2}, contours of constant energy density at time $\tau$=8.6 fm is shown. Here also we find than at any point
energy density of viscous fluid is higher than its ideal counter part. 
The result is in accordance with our
expectation. For dissipative fluid, equation of motion can be written as,

\begin{equation}
\label{eq61a}
D\varepsilon=-(\varepsilon+p)\nabla_\mu u^\mu + \pi^{\mu\nu} \nabla_{<\mu} u_{\nu>}\\
\end{equation}

Due to viscosity, evolution of energy density (or temperature) is slowed down.

In Fig.\ref{F3} and \ref{F4}, we have shown the contour plot of the fluid velocity, $v_x$ and $v_y$, after evolution of 8.6 fm.
As before the black lines
are for the ideal fluid evolution. The red and blue lines are for viscous fluid
with $\eta/s$=0.08 and 0.135 respectively.  Fluid velocities in viscous and ideal fluid differ very little. Even at late time, as shown in Fig.\ref{F3} and \ref{F3}, we find that for $\eta/s$=0.08-0.135, x and y component of the fluid velocity show marginal difference.  However, there is an indication that  in a viscous fluid, velocity grow faster than in ideal fluid. But as mentioned earlier, the difference is marginal.

\begin{figure}[th]
\includegraphics[bb=14 13 581 829,width=0.9\linewidth,clip]{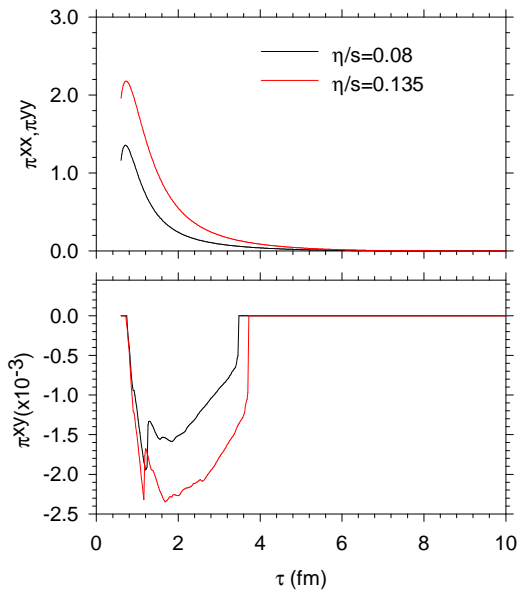}
\vspace{-2.5cm}
\caption{(color online). In the upper panel, temporal evolution of the shear pressure tensor $\pi^{xx}$ at the fluid cell x=y=0 is shown. In the lower panel, evolution of $\pi^{xy}$ at the fluid cell x=y=5 fm is shown. The black and red lines are for ADS/CFT motivated viscosity $\eta/s$=0.08 and perturbative estimate $\eta/s$=0.135 respectively.}
\label{F7}
\end{figure}

As seen in Fig.\ref{F1}-\ref{F2}, in viscous dynamics, QGP fluid evolves slowly. Thus life-time of the QGP phase is enhanced in viscous dynamics. 
To obtain an idea about the enhanced life-time,   in Fig.\ref{F5}, we have shown the constant temperature
contours in $\tau-x$ plane ,  at a fixed value of y=0 fm. 
As seen in Fig.\ref{F5}, temperature evolves slowly in a viscous fluid and life-time of the QGP phase is extended. For small viscosity $\eta/s$=0.08-0.135, the increase is not large. At the center of the fluid, for $\eta/s$=0.135, QGP life-time is increased approximately by 5\% only. It is even less for the ADS/CFT estimate of viscosity. However, enhancement of QGP life-time depends on the fluid cell position. It could be more.  
In Fig.\ref{F6}, constant temperature contours at y=5 fm is shown. For $\eta/s$=0.135, at x=0,y=5 fm, the QGP life-time is enhanced by $\sim$ 10\%. We conclude that in a viscous dynamics, with moderate viscosity $\eta/s$=0.08-0.135, QGP life-time could be enhanced by 5-10\%. Enhanced lifetime of QGP in a viscous fluid can have significant effect on 
observables produced early in the collisions e.g. direct photon
production or in $J/\psi$ suppression.
 
\begin{figure}[th]
\includegraphics[bb=14 13 581 829,width=0.9\linewidth,clip]{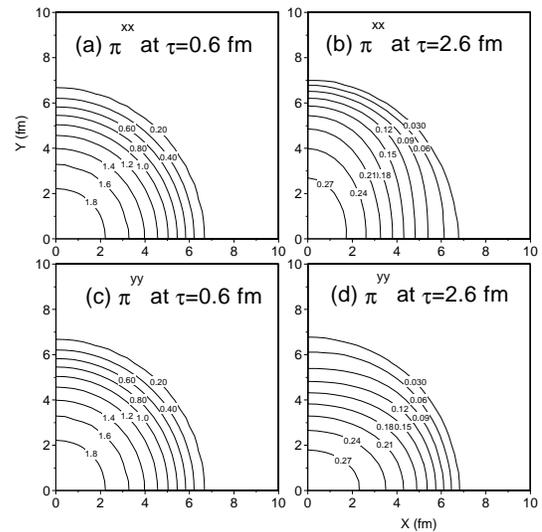}
\vspace{-3cm}
\caption{(color online). In panel (a) and (b), contours of constant pressure tensor $\pi^{xx}$ at initial time $\tau_i$=0.6 fm and
at time $\tau$=2.6 fm is shown. In panel (c) and (d) same results for shear pressure tensor $\pi^{yy}$ is shown.}
\label{F8}
\end{figure}

\subsection{Evolution of shear pressure tensors}

We have assumed that initially the  shear pressure tensors $\pi^{xx}$, $\pi^{yy}$ and $\pi^{xy}$ attained their longitudinal boost-invariant  values. As the fluid evolve, pressure tensors also evolve.  Here we investigate the evolution of shear pressure
tensors with time.
In the top panel of Fig.\ref{F7} evolution
of shear pressure tensor $\pi^{xx}$  at the fluid cell position x=y=0 is shown. The black line is for the ADS/CFT motivated viscosity, $\eta/s$=0.08 and the red line is for the perturbative estimate of viscosity $\eta/s$=0.135.
Just after the start of the evolution the shear pressure  tensor $\pi^{xx}$ increases, but for a short duration and then 
steadily decreases with time. By 4 fm of evolution, $\pi^{xx}$ at the center of the fluid reduces to negligibly small values. Identical behavior is seen for the shear pressure tensor $\pi^{yy}$.   In the bottom panel of Fig.\ref{F7} we have shown
the evolution of the third independent shear pressure tensor $\pi^{xy}$. Initially $\pi^{xy}$ is zero. As the fluid evolve, it grow in the negative direction. We find that at the centre of the fluid (x=y=0), it never grows. In Fig.\ref{F7}, temporal evolution of $\pi^{xy}$  at the fluid cell position $x=y=5 fm$ is shown.
 From the initial zero value, $\pi^{xy}$ rapidly increases in the negative direction. It reaches its  maximum around $\tau \approx$1 fm and  then decreases again. We also note that $\pi^{xy}$ never grows to large values. Compared to $\pi^{xx}$ or $\pi^{yy}$ stress tensor $\pi^{xy}$ is negligible.
The results indicate that in a QGP fluid, viscous effect persist
for a short duration (3-4 fm) only. At late time the fluid evolve essentially as an ideal fluid. The result is understandable.
Shear viscosity depend strongly on temperature ($\eta \propto T^3$). As the fluid cools, effect of viscosity decreases rapidly.

\begin{figure}[ht]
\includegraphics[bb=14 13 581 829,width=0.9\linewidth,clip]{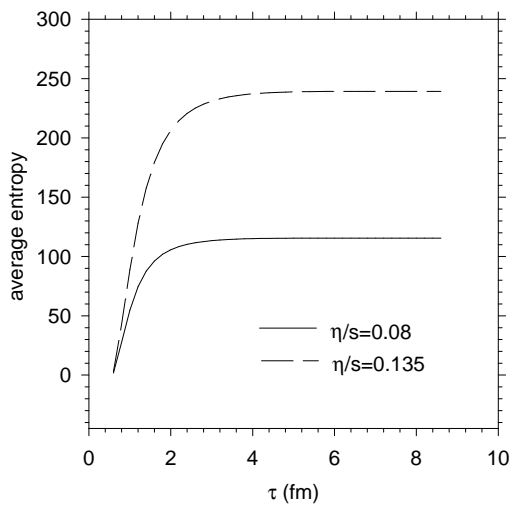}
\vspace{-3.5cm}
\caption{  Evolution of average entropy with time,
for two values of viscosity, the 
  ADS/CFT motivated viscosity $\eta/s$=0.08 and perturbative estimate $\eta/s$=0.135 are shown.}
\label{F9}
\end{figure}

To show the spatial distribution of the stress tensors, in Fig.\ref{F8}, $\pi^{xx}$ and $\pi^{yy}$ at initial time $\tau_i$=0.6 fm and after an evolution of $\tau=$ 2.6 fm are shown. As shown earlier, $\pi^{xx}$ and also $\pi^{yy}$ rapidly decreases with time. By 2 fm of evolution they are reduced by approximately by a factor 6.
It is also interesting to note that the initial x-y symmetric distribution of $\pi^{xx}$ and $\pi^{yy}$ quickly evolves to asymmetric distribution. With time
$\pi^{xx}$ evolves faster in the x-direction than in y-direction. Similarly, $\pi^{yy}$ evolve faster in the y-direction than in the 
x-direction. For central collisions the asymmetric evolution of
$\pi^{xx}$ and $\pi^{yy}$ counter balance each other. As shown
in Fig.\ref{F1} and \ref{F2}, the contour plots of energy density
do not show any indication of asymmetry even at late time.
However, the asymmetric pressure tensors can have important effects on elliptic flow of observables produced early in the collisions, say in elliptic flow of direct photons. 

\subsection{Entropy generation} 

In a viscous fluid dynamics, entropy is generated. We can easily
calculate the entropy generated during the evolution,
 
\begin{eqnarray}
\partial_\mu S^\mu &=& \frac{\pi^{\mu\nu}\pi_{\mu\nu}}{2\eta T} \nonumber \\
&=& \frac{1}{2\eta T} [(\pi^{\tau\tau})^2+(\pi^{xx})^2+(\pi^{yy})^2+(\tau^2\pi^{\eta\eta})^2 \nonumber \\
&& -2(\pi^{\tau x})^2 -2(\pi^{\tau y})^2 
 + 2(\pi^{xy})^2  ]
\end{eqnarray}

Evolution of spatially averaged entropy is shown in Fig.\ref{F9},
for the two values of viscosity coefficients $\eta/s$=0.08 and 0.135.
As expected, entropy generation is more if viscosity is more.
For both the values of viscosity, we find that entropy generation saturates after $\approx$ 3 fm of evolution. It is expected also. As shown
previously, viscous fluxes reduces to very small values after $\tau$=3 fm. Naturally, entropy generation is negligible thereafter.

\begin{figure}[ht]
\includegraphics[bb=14 13 581 829,width=0.99\linewidth,clip]{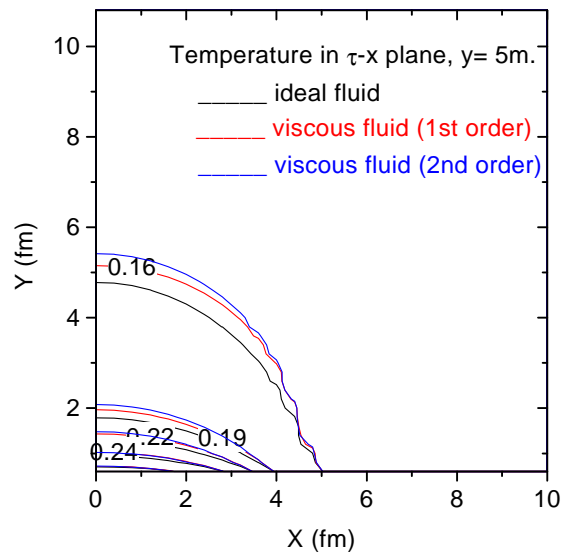}
\vspace{-4.5cm}
\caption{  (color online) constant temperature contours in $x-\tau$ plane at y=5 fm. The black lines are for ideal fluid. The red and blue lines are for viscous fluid in 1st order and in 2nd order theory respectively. $\eta/s$=0.135.
}
\label{F10}
\end{figure}

\subsection{1st order theory vs. 2nd order theory} 

\begin{figure}
\includegraphics[bb=14 13 581 829,width=0.9\linewidth,clip]{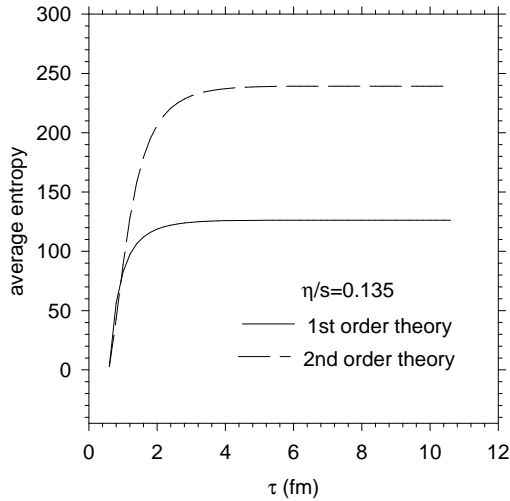}
\vspace{-4cm}
\caption{ Evolution of average entropy production in a 1st order (solid line)
and 2nd order (dashed line) theory. 2nd order theory generate more entropy.}
\label{F11}
\end{figure}

As mentioned earlier, 1st order theory of dissipative hydrodynamics is acausal, signal can travel faster than light. This is corrected in 2nd order theory, but we have to pay the price, relaxation equations for dissipative fluxes are required to be solved.  It is interesting to
compare the difference we can expect in a first order theory and in a 2nd order theory of dissipation.  In Fig. \ref{F10}, we have shown the contours of constant temperature in $x-\tau$, for a fixed $y=5 fm$. The black lines are for an ideal fluid. The red lines
are for a viscous fluid treated in the 1st order theory. The blue lines
are for viscous fluid in 2nd order theory. In 2nd order theory fluid
evolve more slowly than in a first order theory. Entropy generation is also more in a 2nd order theory. In Fig.\ref{F11}, average
entropy evolution with proper time is shown, both for the
1st order theory (the solid line) and the 2nd order theory.
In 2nd order theory, approximately 80\% more entropy is 
generated.

\section{Transverse momentum and elliptic flow of quarks}\label{sec6}

Presently we can not compare  predictions from viscous hydrodynamics with experimental data. Hadrons are not included in the model. The initial QGP fluid evolve and cools
but remain in the QGP phase, it did not undergo a phase transition to hadronic gas. However, from the momentum distribution of quarks we can get some idea about the viscous effect on particle
production. Viscosity generates entropy, which will be reflected in enhanced multiplicity. We use the  standard Cooper-Frey 
prescription to obtained the transverse momentum distribution of quarks. In Cooper-Frey prescription, particle distribution is obtained by convoluting the one body distribution function over the freeze-out surface,

 \begin{equation} \label{eq6_1}
E\frac{dN}{d^3p}=  =\int_\Sigma d\Sigma_\mu p^\mu f(x,p)
\end{equation}

\noindent where $d\Sigma_\mu$ is the freeze-out hyper-surface and $f(x,p)$ is the
one-body distribution function. Now in a viscous dynamics, the fluid is not in
 equilibrium and $f(x,p)$ can not be approximated by the equilibrium distribution function,
 
\begin{equation}\label{eq6_2}
f^{(0)}(x,p)=\frac{1}{exp[\beta(u_\mu p^\mu -\mu)] \pm 1},
\end{equation} 

\noindent with inverse temperature $\beta=1/T$ and chemical potential
$\mu$.
  In a highly
non-equilibrium system,   distribution
function $f(x,p)$ is unknown. 
If the system is slightly off-equilibrium,  then
it is possible to calculate correction to equilibrium distribution 
function due to   (small) non-equilibrium effects. Slightly
off-equilibrium distribution function can be  approximated  as,

\begin{equation} \label{eq6_3}
f(x,p)=f^{(0)}(x,p) [1+\phi(x,p)],
\end{equation}

\noindent  $\phi(x,p)$ is the deviation from equilibrium distribution 
function  $f^{(0)}$. With shear viscosity as the only dissipative forces,
$\phi(x,p)$ can be locally approximated by a quadratic function 
of 4-momentum,

\begin{equation}
\phi(x,p)=\varepsilon_{\mu\nu} p^\mu p^\nu.
\end{equation}

Without any loss of generality $\varepsilon_{\mu\nu}$ can be written as
as,

\begin{equation} \label{eq6_5}
\varepsilon_{\mu\nu}=\frac{1}{2(\varepsilon+p)T^2} \pi^{\mu\nu},
\end{equation}

\noindent completely specifying the non-equilibrium distribution function. As expected, correction factor increases with increasing viscosity. We also note that non-equilibrium correction is more on large momentum particles. 
The effect of viscosity is more on large momentum particles. The correction factor reduces if 
freeze-out occurs at higher temperature.  

With the corrected distribution function, we can calculate the quark momentum
spectra at freeze-out surface $\Sigma_\mu$. In appendix \ref{app5}, relevant
equations are given. The quark momentum distribution has two parts,
(i) $\frac{dN^{eq}}{dyd^2p_T}$, obtained by convoluting the equilibrium distribution function
over the freeze-out surface and (ii)$\frac{dN^{neq}}{dyd^2p_T}$, obtained by convoluting the
correction to the equilibrium distribution function over the freeze-out surface. 
Since the correction factor is obtained under the assumption that non-equilibrium effects are small, $\phi(x,p) << 1$, it necessarily imply that, 
$\frac{dN^{neq}}{dyd^2p_T} << \frac{dN^{eq}}{dyd^2p_T}$. The ratio,
  
  \begin{figure}
\includegraphics[bb=14 13 581 829,width=0.9\linewidth,clip]{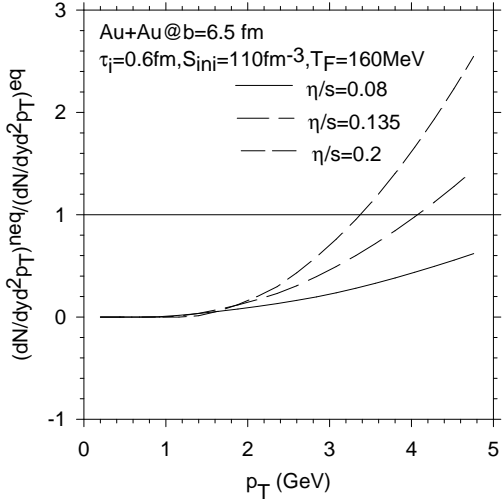}
\vspace{-4cm}
\caption{ Ratio of quark spectra with non-equilibrium distribution function to that with equilibrium distribution function.} 
\label{F12}
\end{figure}
  
\begin{equation}
\frac{dN^{neq}}{dN^{eq}}=\frac
{\frac{dN^{neq}}{dyd^2p_T}}{\frac{dN^{eq}}{dyd^2p_T}},
\end{equation}

\noindent could at best be unity or less. If the ratio exceeds unity, it will imply 
that non-equilibrium effects are large and the distribution function $f(x,p)$ can not be approximated as in Eq.\ref{eq6_3}. Using AZHYDRO-KOLKATA, we have simulated a b=6.5 fm Au+Au collision.
$\frac{dN^{eq}}{dyd^2p_T}$ and $\frac{dN^{neq}}{dyd^2p_T}$ at freeze-out temperature $T_F$=160 MeV are calculated. The ratio $\frac{dN^{neq}}{dN^{eq}}$ for $\eta/s$=0.08,0.135 and 0.2, are shown in    Fig.\ref{F12}. With ADS/CFT estimate of viscosity, 
$\eta/s$=0.08, non-equilibrium correction to particle 
production become comparable to equilibrium contribution only beyond 
$p_T$=5 GeV. However, with perturbative estimate, $\eta/s$=0.135,
non-equilibrium correction become comparable to or exceeds the equilibrium contribution at $p_T\sim$ 4 GeV. $p_T$ range is further reduced for higher
viscosity $\eta/s$=0.2. 
 Thus with perturbative
estimate of viscosity ($\eta/s=0.135-0.2)$, hydrodynamic description
remain valid upto transverse momentum $p_T \sim$ 3.5-4 GeV.

\begin{figure} 
\includegraphics[bb=14 13 581 829,width=0.9\linewidth,clip]{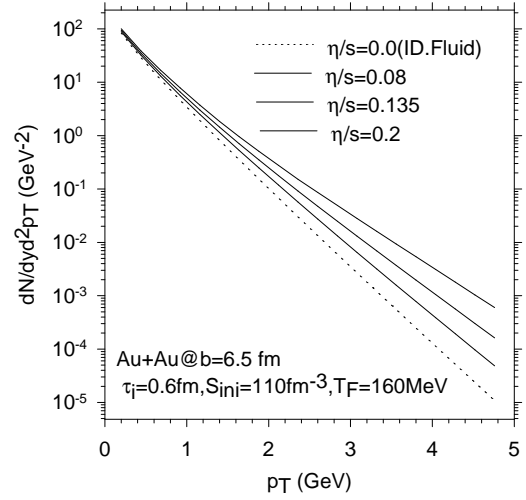}
\vspace{-4cm}
\caption{Qurak transverse momentum spectra  at freeze-out temperature of 160 MeV. The dotted line is the  quarks spectra
in ideal hydrodynamics. The solid lines (top to bottom) are in viscous dynamics with $\eta/s$=0.08,0.135 and 0.2.} 
\label{F13}
\end{figure}

In Fig.\ref{F13}, we have compared the transverse momentum spectra of quarks in ideal hydrodynamics with that in a viscous dynamics. In Fig.\ref{F13}, the dotted line is the spectra obtained in ideal dynamics ($\eta/s=0$). The $p_T$ spectra in viscous 
dynamics are shown by black lines. We have shown the spectra for three values of viscosity $\eta/s$=0.08,0.135 and 0.2.   Compared to ideal dynamics, quarks yield in viscous dynamics increases. The increase is more at large $p_T$. For low values of viscosity the increase is modest,
a factor of 2 at $p_T$=3 GeV. But yield increase by a factor or 4(10) if viscosity increases to $\eta/s$=0.135 (2).  Please
note that even though we have shown $p_T$ spectra upto 5 GeV, for $\eta/s$=0.2 and 0.135, hydrodynamic description fails beyond
$p_T\sim$ 3.5 and 4 GeV.
 
\begin{figure}
\includegraphics[bb=32 290 524 771,width=0.9\linewidth,clip]{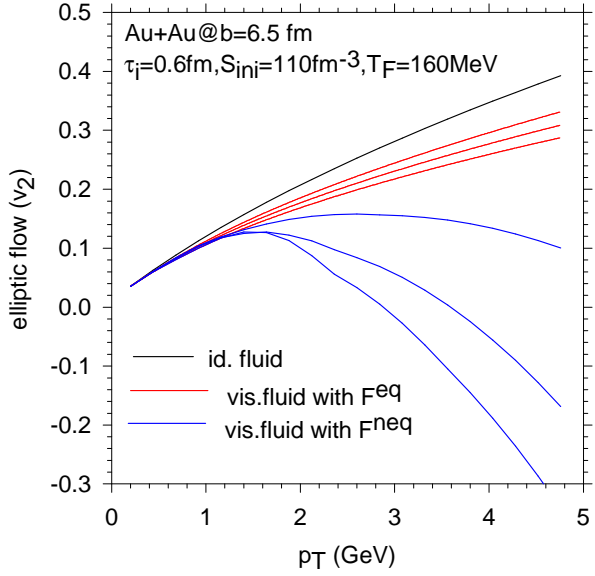}
\caption{(color online) Elliptic flow as a function of transverse momentum. The black line is $v_2$ in ideal hydrodynamics. 
The blue lines are $v_2$ in viscous dynamics with viscosity to entropy ratio $\eta/s$=0.08,0.135 and 0.2 (top to bottom) respectively, including the correction to equilibrium distribution function.  
The red lines are same as the blues lines but ignoring the non-equilibrium correction to the distribution function.} 
\label{F14}
\end{figure}

We have also studied the effect of viscosity on quark elliptic flow. 
Effect of viscosity is very prominent on elliptic flow.  
In Fig.\ref{F14}, $p_T$ dependence of elliptic flow of quarks, in a b=6.5 fm collision is shown.
The black line is $v_2$ in ideal dynamics. In ideal dynamics, elliptic flow continually increases with $p_T$. It well known, in contrast to experiments, where elliptic flow saturates at large $p_T$, in ideal hydrodynamics, elliptic flow continue to increase with $p_T$. Indeed, this is a major problem in ideal hydrodynamics. The renewed the interest in dissipative hydrodynamics is partly due to the inability of ideal hydrodynamics to predict the trend of elliptic flow in Au+Au collisions.   In Fig.\ref{F14},
the blue lines are $v_2$  in  viscous dynamics with $\eta/s$=0.08,0.135 and 0.2 respectively. In a viscous dynamics,  $p_T$ dependence
of $v_2$ is drastically changed. In contrast to ideal dynamics where $v_2$ continue to increase with $p_T$, in
 viscous dynamics,
$v_2$ continue to increase only upto $p_T\sim 1.5-2 GeV$. Thereafter $v_2$ decreases. For perturbative estimate of viscosity $\eta/s$=0.135 and beyond, $v_2$ even become negative at large $p_T$ . Veering about of $v_2$ after $p_T\sim $1.5-2 GeV is due to viscous effect only or more explicitly due to the non-equilibrium correction to the equilibrium distribution function. This is clearly
manifested from the red lines in Fig.\ref{F14}. The red lines are calculated ignoring the non-equilibrium corrections to the equilibrium distribution function. If non-equilibrium correction is ignored, in viscous dynamics also, $v_2$ continue to increase with $p_T$,    albeit its magnitude is reduced compared to ideal dynamics.  The result is very important. It imply that the experimental trend of elliptic flow (saturation at large $p_T$) could
only be explained if the QGP fluid is viscous. An ideal QGP, will not be  able to explain the saturation trend of the experimental data. 
 
\begin{figure}
\includegraphics[bb=38 316 534 795,width=0.9\linewidth,clip]{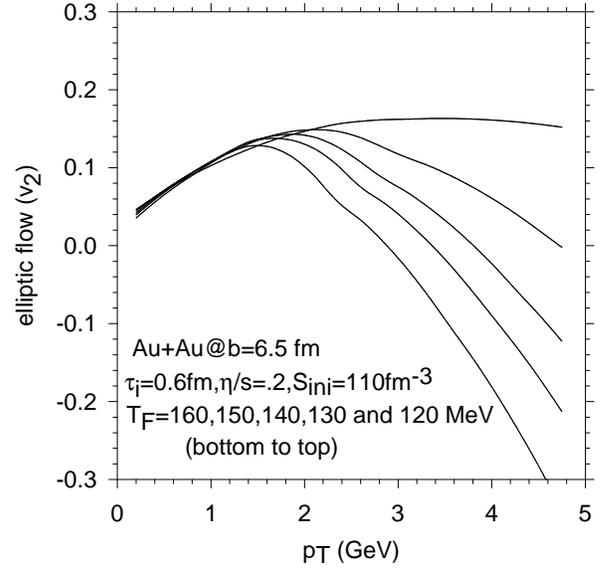}
\caption{ Dependence of elliptic flow on the freeze-out temperature. The solid lines (from bottom to top) are elliptic flow
($v_2$) in viscous dynamics with 
$T_F$=160,150,140,130 and 120 MeV respectively.  } 
\label{F15}
\end{figure}

As stated earlier, non-equilibrium correction to the equilibrium distribution function depends   on the freeze-out condition. To show the effect of freeze-out condition, on $v_2$,
in Fig.\ref{F15} we have shown $v_2$ for a values of freeze-out temperature $T_F$ =160,150,140,130 and 120 MeV. As freeze-out occur at higher and higher temperature, the veering of $v_2$ takes place at larger and larger $p_T$ and for $T_F$=120 MeV, the elliptic flow  saturates. The result is understood easily. With decreasing freeze-out temperature, the fluid evolves for longer time, the shear stress-tensor's at the freeze-out surface is reduced and the non-equilibrium correction, proportional to shear stress tensors, decreases.

\section{Stability of numerical solutions in AZHYDRO-KOLKATA} 

\begin{figure}
\includegraphics[bb=21 308 507 796,width=0.9\linewidth,clip]{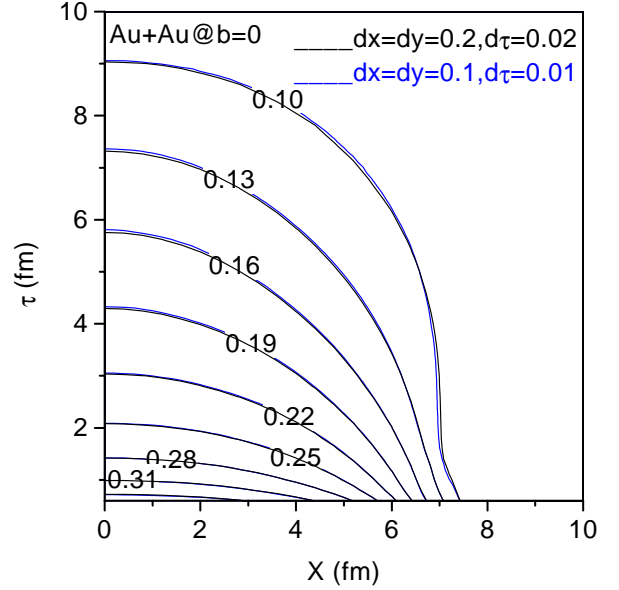}
\caption{ (color online) constant temperature contours in $x-\tau$ plane at a fixed y=0 fm. The black lines are obtained with integration step lengths dx=dy=0.2 fm and $d\tau$=0.02 fm. The blue lines are obtained with integration step lengths,
dx=dy=0.1 fm and $d\tau$=0.01 fm. Halving the step lengths do not change the evolution. The numerical solutions are stable.}
\label{F16}
\end{figure}

Before we summarise our results, we would like to comment on the
stability of numerical solutions in AZHYDRO-KOLKATA.
As indicated above, with shear viscosity as the only dissipative force, boost-invariant causal hydrodynamics require simultaneous solution of six partial differential equations. Numerical solution of six partial differential equations is non-trivial and it is important to check for the numerical stability of the solutions. 
Analytical solutions of viscous hydrodynamics, even in restrictive conditions are not available, and we can not check the solutions against  analytical results. However, we can check for the stability of the numerical solutions. The standard procedure of checking the numerical stability is to change the integration step lengths and look for the difference in the solution.  
 In Fig.\ref{F16}, for viscosity $\eta/s$=0.135, we have shown the constant temperature contours in $x-\tau$ plane at a fixed y=0 fm. The black and blue lines are obtained when integration step lengths are dx=dy=0.2fm,$d\tau$=0.02 fm,
and dx=dy=0.1 fm,$d\tau$=0.01fm respectively. Evolution
of QGP fluid donot alter by changing the step lengths, the solutions are stable against mesh size. 

\section{Summary and conclusions}
\label{sec7}

In Israel-Stewart's 2nd order theory of dissipative relativistic hydrodynamics, we have studied evolution QGP fluid. In 2nd order theory, in addition to usual thermodynamic quantities e.g. energy density, pressure, hydrodynamic velocities, dissipative flows are treated as extended thermodynamic variables. Relaxation equations for dissipative
flows are   solved, simultaneously with the energy-momentum conservation equations. This greatly enhances the complexity of the problem. Altogether  14 partial differential
equations are required to be solved. We simplify the problem
to solution of six partial differential equations
by considering the evolution of baryon free QGP fluid with longitudinal boost-invariance. We also consider 
dissipation due to shear viscosity only, disregarding the bulk viscosity and the heat conduction (for a baryon free QGP fluid they
do not contribute). The six partial differential equations are solved using the code AZHYDRO-KOLKATA, developed at the Cyclotron Centre, Kolkata. 

To bring out the effect of viscosity, we have considered the evolution of ideal as well as viscous QGP fluid. Both ideal
and viscous fluid are initialized similarly, at initial time $\tau_i$=0.6 fm, the central entropy density is 110 $fm^{-3}$.
Viscous dynamics require initial conditions  for the shear-stress tensor components. It is assumed that at the equilibration time, the shear stress tensors components have attained their boost-invariant values.  

Explicit simulation of ideal and viscous fluids confirms that energy density of a 
viscous fluid, evolve slowly than its ideal counterpart. Thus
in a viscous fluid, lifetime of the QGP phase will be enhanced.
Transverse expansion is also more in viscous dynamics.
For a similar freeze-out condition freeze-out surface is extended in
viscous fluid. 

As the fluid evolve, shear pressure tensors also evolve. Explicit simulations indicate that shear pressure tensors $\pi^{xx}$ and $\pi^{yy}$ which are initially non zero, rapidly decreases as the fluid evolve. By 3-4 fm of evolution they reduced to very small values. The other independent shear tensor $\pi^{xx}$ is zero 
initially. At later time it grow in the negative direction but 
 never grow to large value and is always order of magnitude smaller than the stress tensors ($\pi^{xx}$ and $\pi^{yy}$).  
  Spatial distribution of shear pressure tensors $\pi^{xx}$ and $\pi^{yy}$ reveal an interesting feature of viscous dynamics. Initially $\pi^{xx}$ and $\pi^{yy}$  have symmetric distribution. As the fluid evolve, pressure tensors quickly become asymmetric, e.g. $\pi^{xx}$ evolve faster in the
x-direction than in the y-direction, $\pi^{yy}$ evolve faster in y direction than in x-direction. However, in a central collision, we did not see any effect of asymmetry in the energy density distribution. In a central
b=0 collision, the two opposite asymmetry cancels each other.  
 
We could not study effect of shear viscosity on particle production.
However, we have explored the effect of viscosity on parton momentum distribution and elliptic flow. We have simulated 
b=6.5 fm Au+Au collision. Using the Cooper-Frey prescription, transverse momentum spectra as well as elliptic flow of quarks at freeze-out temperature of $T_F$=160 MeV are obtained. 
Viscous dynamics flattens the quark yield at large $p_T$.
At $p_T$=3 GeV, even a small viscosity, $\eta/s$=0.8, increase the yield by a factor of 2. The increase is even more if viscosity is large. Viscous effect is most prominent on elliptic flow. 
 In ideal hydrodynamics, elliptic flow continue to increase with $p_T$. But in viscous dynamics  $v_2$ veer about around $p_T$=1.5-2 and even become negative at large $p_T$. With appropriate choice of viscosity, freeze-out condition, elliptic flow show saturation.
The saturation effect is essentially due to non-equilibrium correction to the equilibrium distribution function and can not be mimicked in an ideal hydrodynamics. Only in viscous dynamics,  saturation of elliptic flow can be explained.

\appendix
\section{Coordinate transformations}\label{app1}

Instead of Cartesian coordinates $x^\mu=(t,x,y,z)$ we use curvilinear coordinates in longitudinal proper time and rapidity,
$\bar{x}^m=(\tau,x,y,\eta)$:
\begin{eqnarray}
t &=& \tau \cosh\eta ; \hspace{1.6cm} \tau=\sqrt{t^2-z^2} \\
z &=& \tau \sinh\eta; \hspace{1.6cm} \eta=\frac{1}{2}\ln \frac{t+z}{t-z}.
\end{eqnarray}
The differentials 
\begin{eqnarray}
dt &=& d\tau \cosh\eta + d\eta\, \tau \sinh\eta, \\ 
dz &=& d\tau \sinh\eta+ d\eta\, \tau \cosh\eta,
\end{eqnarray}
and the metric tensor is easily read off from
\begin{eqnarray}
  ds^2 &=& g_{\mu\nu} dx^\mu dx^\nu = dt^2-dx^2-dy^2-dz^2 
\nonumber\\
       &=& \bar g_{mn} d\bar x^m d\bar x^n =
           d\tau^2 -dx^2 -y^2 -\tau^2 d\eta^2,
\end{eqnarray}
namely
\begin{equation}
\bar g_{mn}=\left(
\begin{array}{cccc}
   1      & 0   & 0   & 0   \\
   0      & -1  & 0   &  0  \\
   0      & 0   &  -1  & 0   \\
   0      &  0  & 0  &  -\tau^2  \\
\end{array} \right), 
\bar g^{mn}=\left(
\begin{array}{cccc}
   1      & 0   & 0   & 0   \\
   0      & -1  & 0   &  0  \\
   0      & 0   &  -1  & 0   \\
   0      &  0  & 0  &  -1/\tau^2  \\
\end{array} \right)
\end{equation}
In curvilinear coordinates we must replace the partial derivatives 
with respect to $x^\mu$ by covariant derivatives (denoted by a 
semicolon) with respect to $\bar x^m$:
\begin{eqnarray*}
{\bar T^{ik}}_{;p} &=&\frac{\partial \bar T^{ik}}{\partial \bar x^p} 
+ \Gamma^i_{pm} \bar T^{mk} + \bar T^{im}\Gamma^k_{mp}.
\end{eqnarray*}
The only non-vanishing Christoffel symbols are
\begin{equation}
\Gamma^\tau_{\eta\eta}=\tau;
\quad
\Gamma^\eta_{\tau\eta}=\Gamma^\eta_{\eta\tau}=1/\tau.
\quad
\end{equation}
The hydrodynamic 4-velocity $u^\mu=\gamma(1,v_x,v_y,v_z)$ is transformed 
to $\bar u^m=\gamma(1,v_x,v_y,0)$, with $\gamma_\perp=1/\sqrt{1{-}v^2_r}$.
From here on, we drop the bars over tensor components in 
$\bar x$-coordinates for simplicity.

The projector can be easily calculated,
\begin{eqnarray}
\label{A16}
\Delta^{\mu\nu}  =&&g^{\mu\nu} - u^\mu u^\nu \nonumber\\
=&&
\left(\begin{array}{cccc}
1-\gamma_\perp^2 & -\gamma_\perp^2 v_x & \gamma_\perp^2 v_y       & 0\\ 
-\gamma_\perp^2 v_x  & -1-\gamma_\perp^2 v_x^2  & -\gamma_\perp^2v_x v_y   & 0\\ 
-\gamma_\perp^2 v_y & -\gamma_\perp^2 v_x v_y    & -1-\gamma_\perp^2 v_y^2 & 0\\
                0      &    0       & 0         & \frac{1}{\tau^2}
      \end{array}
\right).
\end{eqnarray}

In $(\tau,x,y,\eta)$ coordinate system, the convective time derivative can be obtained as,
\begin{equation} \label{eqa9}
D = u\cdot\partial = \gamma(\partial_\tau + v_x \partial_x+v_y \partial_y).
\end{equation}

For future reference,we also write down the 
the scalar expansion rate
\begin{equation} \label{eqa10}
\theta= \partial{\cdot}u 
= \partial_\tau u^\tau + \partial_x u^x +  \partial_y u^y+ \frac{u^\tau}{\tau} 
\end{equation}
%

\section{Energy-momentum conservation} \label{app2}

With longitudinal boost-invariance   the 
energy-momentum conservation equations ${T^{mn}}_{;n}=0$ yield 

 \begin{widetext}
\begin{eqnarray}
\label{eqb1}
&& \partial_\tau \tilde{T}^{\tau\tau}  
 +\partial_x (\tilde{T}^{\tau \tau} \overline{v}_x  )+\partial_y  (\tilde{T}^{\tau \tau} \overline{v}_y)= 
  -\,(p+\tau^2 \pi^{\eta\eta})
\\
\label{eqb2}
&&\partial_\tau\tilde{T}^{\tau x} 
 +\partial_x (\tilde{T}^{\tau x}v_x)
 +\partial_y (\tilde{T}^{\tau x}v_y) 
 = -\partial_x(\tilde{p} + \tilde{\pi}^{xx}-\tilde{\pi}^{\tau x} v_x) - \partial_y(\tilde{\pi}^{xy}-\tilde{\pi}^{\tau x}v_y)
\\
\label{eqb3}
 &&\partial_\tau\tilde{T}^{\tau y} 
 +\partial_x (\tilde{T}^{\tau y}v_x)
 +\partial_y (\tilde{T}^{\tau y}v_y) 
 = -\partial_x(\tilde{\pi}^{xy}-\tilde{\pi}^{\tau y} v_x) - \partial_y(\tilde{p} + \tilde{\pi}^{yy}-\tilde{\pi}^{\tau y}v_y)
\end{eqnarray}
\end{widetext}

\noindent where $\tilde{A}^{mn}\equiv  \tau A^{mn}$,
$\tilde{p}\equiv \tau p$, and 
$\overline{v}_x \equiv T^{\tau x}/T^{\tau\tau}$,
$\overline{v}_y \equiv T^{\tau y}/T^{\tau\tau}$.

The components of the energy momentum tensors, including the shear pressure tensor are, 

\begin{eqnarray}
\label{eqb4}
T^{\tau\tau} =&& (\varepsilon+p)\gamma_\perp^2 - p + \pi^{\tau\tau}\\
\label{eqb5}
T^{\tau x} =&& (\varepsilon+p)\gamma_\perp^2 v_x + \pi^{\tau x}\\
\label{eqb6}
T^{\tau y} =&& (\varepsilon+p)\gamma_\perp^2 v_y + \pi^{\tau y}
\end{eqnarray}

In causal dissipative hydrodynamics, energy momentum conservation equations are solved simultaneously with the
relaxation equations.
Given an equation of state, if
energy density ($\varepsilon$) and fluid velocity ($v_x$ and $v_y$) distributions, at any time
$\tau_i$ are known, Eqs.\ref{eqb1},\ref{eqb2} and \ref{eqb3} can be  integrated
to obtain $\varepsilon$, $v_x$ and $v_y$ at the next time step $\tau_{i+1}$. 
While for ideal hydrodynamics, this procedure works perfectly,   
  viscous hydrodynamics poses a problem
that shear stress-tensor components contains 
time derivatives, $\partial_\tau \gamma_\perp$, $\partial_\tau u^x$,
$\partial_\tau u^x$ etc. Thus at time step $\tau_i$ one needs the  
still unknown time derivatives.  
Numerically, time derivatives at step $\tau_i$ could be obtained
if velocities at time step $\tau_i$ and $\tau_{i+1}$ are known.
One possible way to circumvent the problem, is to use time derivatives of the previous step, i.e. use velocities at time step $\tau_{i-1}$ and $\tau_i$ to calculate the derivatives at time step $\tau_i$ \cite{Chaudhuri:2005ea}. The underlying assumption that fluid velocity
changes slowly with time.  
In 1st order theories, this problem is circumvented by calculating the time derivatives from the ideal equation of motion ,

\begin{eqnarray}
\label{eqb7}
Du^\mu &=&\frac{\nabla^\mu p}{\varepsilon+p},\\
\label{eqb8}
D\varepsilon&=&-(\varepsilon+p)\nabla_\mu u^\mu.
\end{eqnarray}

With the help of these two equations all the time derivatives can be 
expressed entirely in terms of spatial gradients \cite{Teaney:2004qa,deGroot}. 
1st order theories are restricted to contain terms at most linear
in dissipative quantities.  Neglect of viscous terms can contribute only
in 2nd order corrections, which are neglected in 1st order theories.
While the procedure is not correct in 2nd order theory, we still use it in the present calculations. The alternative procedure of using the
derivative of earlier time step is not correct either. 

\section{Relaxation equations for the viscous pressure tensor} \label{app3}

Being symmetric and traceless, the viscous pressure tensor $\pi^{\mu\nu}$ has 9 independent components. The assumption of boost invariance reduces this number by 3 
($\nabla^{\left\langle m\right.}u^{\left.\eta\right\rangle}{\,=\,}0, \ 
m \neq \eta$).   The transversality condition $u_m \pi^{mn}=0$
eliminates another three components (  $u_\eta$ vanish and 
thus yield no constraint). Thus, with boost-invariance   the viscous pressure tensor has only three independent components.
As seen in Eqs.\ref{eqb1},\ref{eqb2} and \ref{eqb3} in a boost-invariant evolution only seven pressure
tensors $\pi^{\tau\tau}$, $\pi^{xx}$, $\pi^{yy}$, $\pi^{\eta\eta}$,
$\pi^{\tau x}$, $\pi^{\tau y}$ and $\pi^{xy}$ are of importance.
Only three of these seven are independent. 
In an earlier publication \cite{Heinz:2005bw}, we have debated about the choice of 
the independent components and suggested use of either
($\pi^{\tau\tau}$, $\pi^{\eta\eta}$, $\Delta=\pi^{xx}-\pi^{yy}$) or
($\pi^{\tau\tau}$,$\pi^{\eta\eta}$,$\pi^{\tau x}$, $\pi^{\tau y}$) (which will require solution of an additional relaxation equation)
as choice of independent components.
However, while computing we find that the three pressure tensors 
$\pi^{xx}$ and $\pi^{yy}$ and $\pi^{xy}$ as independent components are computationally more convenient. The  choice has the advantage that the dependent shear stress tensors can be obtained
from the 3 independent stress tensors by multiplying them by fluid velocity, $v_x$ and $v_y$ (see Eqs. \ref{eqc1}-\ref{eqc4}).
In any other choice of independent components (e.g. $\pi^{\tau\tau}$,$\pi^{\eta\eta}$,$\Delta=\pi^{xx}-\pi^{yy}$), the evaluation of dependent stress tensors requires division by fluid velocities. Since initially, fluid velocities are assumed to be zero and they grow slowly, these choices will involve  division by very small numbers. Unless proper care is not taken, division by small numbers can lead to unrealistically large values for the dependent stress tensors and ruin the computation.

The relaxation equations for the independent shear stress tensors $\pi^{xx}$, $\pi^{yy}$ and $\pi^{xy}$, in ($\tau$,x,y,$\eta$) co-ordinate can be written as,

\begin{widetext}
\begin{eqnarray}
\label{eqc4a}
\partial_\tau \pi^{xx} +v_x \partial_x \pi^{xx}+v_y \partial_y \pi^{xx}
&=&
-\frac{1}{\tau_\pi \gamma} \left (\pi^{xx} - 2\eta \sigma^{xx}\right )\\
\label{eqc5}
\partial_\tau \pi^{yy} +v_x \partial_x \pi^{yy}+v_y \partial_y \pi^{yy}
&=&
-\frac{1}{\tau_\pi \gamma} \left (\pi^{yy} - 2\eta \sigma^{yy}\right )\\
\label{eqc6}
\partial_\tau \pi^{xy} +v_x \partial_x \pi^{xy}+v_y \partial_y \pi^{xy}
&=&
-\frac{1}{\tau_\pi \gamma} \left (\pi^{xy} - 2\eta \sigma^{xy}\right )
\end{eqnarray}
\end{widetext} 

where $\tau_\pi$ is the relaxation time, $\tau_\pi=2\eta \beta_2$ (see
Eq.\ref{eq21}). In ultra-relativistic limit, for a Boltzman gas, 
$\beta_2$  can be evaluated, $\beta_2 \approx \frac{3}{4p}$ where
$p$ is the pressure \cite{IS79}. In the present paper, we use this limit to
obtain the relaxation time $\tau_\pi$.

The viscous pressure tensor relaxes on a   time scale $\tau_\pi$
to $2\eta$ times the shear tensor $\sigma^{\mu\nu}=
\nabla^{\left\langle\mu\right.}u^{\left.\nu\right\rangle}$.
The $xx$, $yy$  and $xy$ components of the shear tensor $\sigma^{\mu\nu}$ can be written as
%
\begin{eqnarray}
\label{eqc7}
\sigma^{xx}
=&&-\partial_x u^x -u^x Du^x -\frac{1}{3} \Delta^{xx} \theta\\ 
\label{eqc8}
\sigma^{yy}
=&&-\partial_y u^y -u^y Du^y -\frac{1}{3} \Delta^{yy} \theta\\ 
\label{eqc9}
 \sigma^{xy}
=&&-\frac{1}{2}[\partial_x u^y - \partial_y u^x
-u^x Du^y - u^y Du^x] \nonumber \\
&& -\frac{1}{3} \Delta^{xy} \theta
\end{eqnarray}

The dependent shear stress tensors can easily be obtained from
the independent ones as,
  
\begin{eqnarray}
\pi^{\tau x}=&& v_x \pi^{xx} + v_y \pi^{xy} \label{eqc1} \\
\pi^{\tau y}=&& v_x \pi^{xy} + v_y \pi^{yy} \label{eqc2}\\
\pi^{\tau \tau}=&& v^2_x \pi^{xx} + v^2_y \pi^{yy}+2v_x v_y \pi^{xy} \label{eqc3}\\
\tau^2 \pi^{\eta\eta}=&&   
-(1-v^2_x)\pi^{xx} - (1-v^2_y)\pi^{yy} \nonumber\\
&&+2v_xv_y\pi^{xy} \label{eqc4}
\end{eqnarray}

The expressions for the convective time derivative $D$ and
expansion scalar $\theta=\partial \dot u$, in ($\tau$,x,y,$\eta$) 
are given in Eqs. \ref{eqa9} and \ref{eqa10}.

\section{particle spectra}
\label{app5}

With the non-equilibrium distribution function thus specified, it can be used to 
calculate the particle spectra from the freeze-out surface. In the standard 
Cooper-Frye
prescription, particle distribution is obtained as,

\begin{equation} \label{eq5_1}
E\frac{dN}{d^3p}=\frac{dN}{dyd^2p_T} =\int_\Sigma d\Sigma_\mu p^\mu f(x,p)
\end{equation}

In $(\tau,x,y,\eta_s)$ coordinate, the freeze-out surface is parameterised as,

\begin{equation}
\Sigma^\mu=(\tau_f(x,y)\cosh \eta_s, x, y, \tau_f(x,y) \sinh \eta_s),
\end{equation}

\noindent and the normal vector on the hyper surface is,

\begin{equation}
d\Sigma_\mu=(\cosh \eta_s, -\frac{\partial \tau_f}{\partial x_f}, 
                        -\frac{\partial \tau_f}{\partial y_f}, -\sinh \eta_s)
\tau_f dx dy d\eta_s
\end{equation}

At the fluid position $(\tau,x,y,\eta_s)$ the particle 4-momenta are parameterised as,

\begin{equation}
p^\mu=(m_T cosh (\eta_s-Y), p^x, p^y, m_T sinh (\eta_s-Y))
\end{equation}

The volume element $p^\mu d\Sigma_\mu$ become,

\begin{equation}
p^\mu d\Sigma_\mu=(m_T cosh(\eta-Y)-\vec{p}_T. \vec{\nabla}_T \tau_f) \tau_f dx dy d\eta
\end{equation}

Equilibrium distribution function involve the term $\frac{p^\mu u_\mu}{T}$ which can be evaluated as,

\begin{equation}
\frac{p^\mu u_\mu}{T}=\frac{\gamma(m_T cosh(\eta-Y) -\vec{v}_T.\vec{p}_T -\mu/\gamma)}{T}
\end{equation}

The non-equilibrium distribution function require the sum
$p^\mu p^\nu \pi_{\mu\nu}$,

\begin{equation}
p_\mu p_\nu \pi^{\mu\nu}=a_1 cosh^2 (\eta-Y) +a_2 cosh(\eta-Y) + a_3
\end{equation}

with
\begin{eqnarray}
a_1=&&m_T^2(\pi^{\tau \tau} +\tau^2 \pi^{\eta \eta})\\
a_2=&&-2m_T(p_x \pi^{\tau x} + p_y \pi^{\tau y})\\
a_3=&&p_x^2 \pi^{xx} +p_y^2 \pi^{yy} +2p_x p_y \pi^{xy} - m_T^2 \tau^2 \pi^{\eta \eta}
\end{eqnarray}

Inserting all the relevant formulas in Eq.\ref{eq5_1} and integrating over 
spatial rapidity one obtains,
 
\begin{equation}
\frac{dN}{dyd^2p_T} =\frac{dN^{eq}}{dyd^2p_T}+\frac{dN^{neq}}{dyd^2p_T}
\end{equation}

with,

\begin{widetext}
\begin{eqnarray} \label{eq5_12}
\frac{dN^{eq}}{dyd^2p_T}=\frac{g}{(2\pi)^3} 
\int dx dy \tau_f [m_T K_1(n\beta) - p_T \vec{\nabla}_T \tau_f K_0(n\beta)]\\
\frac{dN^{neq}}{dyd^2p_T}=\frac{g}{(2\pi)^3} 
\int dx dy \tau_f [m_T\{ {\frac{a_1}{4} K_3(n\beta)+\frac{a_2}{2} K_2(n\beta)+(\frac{3a_1}{4}+a_3)K_1(n\beta)
+\frac{a_2}{2} K_0(n\beta)} \} \nonumber\\
- \vec{p}_T. \vec{\nabla}_T \tau_f \{\frac{a_1}{2} K_2(n\beta)+
 a_2 K_1(n\beta)+( \frac{a_1}{2}+a_3)K_0(n\beta)\}]
\end{eqnarray}
\end{widetext}

\noindent where  $K_0$, $K_1$, $K_2$ and $K_3$ are the modified Bessel functions.  

We will also show results for elliptic flow $v_2$. It is defined as,

\begin{equation}
V_2=\frac
{\int_0^{2\pi} \frac{dN}{dyd^2p_T} \cos(2\phi) d\phi}
{\int_0^{2\pi} \frac{dN}{dyd^2p_T}  d\phi}
\end{equation}

Expanding to the 1st order, elliptic flow as a function of transverse momentum can be obtained as,

\begin{equation}
v_2(p_T)=v_2^{eq}(p_T) \left(1-\frac{\int d\phi \frac{d^2N^{neq}}{p_T dp_Td\phi}}{\int d\phi \frac{d^2N^{eq}}{p_Tdp_Td\phi}}\right)
+\frac{\int d\phi cos(2\phi) \frac{d^2N^{neq}}{p_T dp_T d\phi}}
{\int d\phi \frac{d^2N^{eq}}{p_T dp_T d\phi}}
\end{equation}

where $v_2^{eq}$ is the elliptic flow calculated with the equilibrium distribution $f^{eq}$.



\begin{thebibliography}{99}
\bibitem{BRAHMSwhitepaper} 
BRAHMS Collaboration, I. Arsene {\it et al.},  
Nucl. Phys. A {\bf 757}, 1 (2005). 
 
\bibitem{PHOBOSwhitepaper} 
PHOBOS Collaboration,  B. B. Back {\it et al.},  
Nucl. Phys. A {\bf 757}, 28 (2005). 
 
\bibitem{PHENIXwhitepaper} 
PHENIX Collaboration, K.~Adcox {\it et al.}, 
Nucl. Phys. A {\bf 757} (2005), in press [arXiv:nucl-ex/0410003]. 
  
\bibitem{STARwhitepaper} 
STAR Collaboration, J. Adams {\it et al.}, 
Nucl. Phys. A {\bf 757} (2005), in press [arXiv:nucl-ex/0501009]. 
\bibitem{lattice} 
Karsch F, Laermann E, Petreczky P, Stickan S and Wetzorke I, 
2001 {\it Proccedings of NIC Symposium} (Ed. H. Rollnik and D. Wolf, John 
von Neumann Institute for Computing, J\"ulich, NIC Series, vol.9, 
ISBN 3-00-009055-X, pp.173-82,2002.)
\bibitem{QGP3}
P.~F. Kolb and U. Heinz,
in {\it Quark-Gluon Plasma 3}, edited by R.~C. Hwa and 
X.-N. Wang (World Scientific, Singapore, 2004), p.~634.




\bibitem{Policastro:2001yc}
  G.~Policastro, D.~T.~Son and A.~O.~Starinets,
  Phys.\ Rev.\ Lett.\  {\bf 87}, 081601 (2001)
  [arXiv:hep-th/0104066].

\bibitem{Policastro:2002se}
  G.~Policastro, D.~T.~Son and A.~O.~Starinets,
  JHEP {\bf 0209}, 043 (2002)
  [arXiv:hep-th/0205052].
\bibitem{Heinz:2004ar}
U.~Heinz,
J.\ Phys.\ G {\bf 31}, S717 (2005).

\bibitem{Heinz:2002un}
  U.~W.~Heinz and P.~F.~Kolb,
  arXiv:hep-ph/0204061.


\bibitem{Eckart}
C.~Eckart, Phys. Rev. {\bf 58}, 919 (1940).
\bibitem{LL63}
L.~D.~Landau and E.~M.~Lifshitz, {\it Fluid Mechanics}, Sect. 127,
Pergamon, Oxford, 1963.

\bibitem{IS79}
W. Israel, Ann. Phys. (N.Y.) {\bf 100}, 310 (1976);
W.~Israel and J.~M.~Stewart, Ann. Phys. (N.Y.) {\bf 118}, 349 (1979).

\bibitem{Muronga:2001zk}
  A.~Muronga,
  Phys.\ Rev.\ Lett.\  {\bf 88}, 062302 (2002)
  [Erratum {\it ibid.}\  {\bf 89}, 159901 (2002)];
  and Phys.\ Rev.\ C {\bf 69}, 034903 (2004).

\bibitem{Teaney:2004qa}
  D.~A.~Teaney,
  J.\ Phys.\ G {\bf 30}, S1247 (2004).

\bibitem{MR04}
  A.~Muronga and D.~H.~Rischke,
  nucl-th/0407114\,(v2).
\bibitem{Heinz:2005bw}
  U.~W.~Heinz, H.~Song and A.~K.~Chaudhuri,
  Phys.\ Rev.\  C {\bf 73}, 034904 (2006)
  [arXiv:nucl-th/0510014].
\bibitem{Chaudhuri:2005ea}
  A.~K.~Chaudhuri and U.~W.~Heinz,
  J.\ Phys.\ Conf.\ Ser.\  {\bf 50}, 251 (2006)
  [arXiv:nucl-th/0504022].


\bibitem{Chaudhuri:2006jd}
  A.~K.~Chaudhuri,
  Phys.\ Rev.\  C {\bf 74}, 044904 (2006)
  [arXiv:nucl-th/0604014].

\bibitem{Chaudhuri:2007yn}
  A.~K.~Chaudhuri,
  arXiv:nucl-th/0703029.
\bibitem{Chaudhuri:2007yk}
  A.~K.~Chaudhuri,
  arXiv:nucl-th/0703027.

\bibitem{Koide:2007kw}
  T.~Koide, G.~S.~Denicol, Ph.~Mota and T.~Kodama,
  Phys.\ Rev.\  C {\bf 75}, 034909 (2007).
\bibitem{Baier:2006gy}
  R.~Baier and P.~Romatschke,
  arXiv:nucl-th/0610108.

\bibitem{Arnold:2000dr}
  P.~Arnold, G.~D.~Moore and L.~G.~Yaffe,
  JHEP {\bf 0011}, 001 (2000)
  [arXiv:hep-ph/0010177].

\bibitem{Baym:1990uj}
  G.~Baym, H.~Monien, C.~J.~Pethick and D.~G.~Ravenhall,
  Phys.\ Rev.\ Lett.\  {\bf 64}, 1867 (1990).

\bibitem{deGroot} S. R. de Groot, W. A. van Leeuwen and
Ch. G. van Weert, {\em Relativistic Kinetic Theory} ( North-Holland, Amsterdam, 1980) p.36

\end{thebibliography}
\end{document}